\documentclass[11pt]{article}
\usepackage[margin=1in]{geometry}
\usepackage{amsmath,amssymb,amsthm,microtype}
\usepackage{booktabs,enumitem,float}
\usepackage{tikz}
\usetikzlibrary{arrows.meta,positioning,calc,decorations.pathreplacing}
\usepackage[hidelinks]{hyperref}
\hypersetup{
  pdftitle={Maximum-Weight Matching in Random-Order Streams: A One-Pass (1\unichar{8722}\unichar{949})-Approximation},
  pdfauthor={Alireza Farhadi, Mohammadtaghi Hajiaghayi, Mohammad Mahdavi}}

\newtheorem{theorem}{Theorem}[section]
\newtheorem{lemma}[theorem]{Lemma}
\newtheorem{proposition}[theorem]{Proposition}

\theoremstyle{definition}
\newtheorem{definition}[theorem]{Definition}
\theoremstyle{remark}
\newtheorem{remark}[theorem]{Remark}

\floatstyle{boxed}
\newfloat{algobox}{t}{loa}
\floatname{algobox}{Algorithm}

\newcommand{\Ball}{\mathrm{Ball}}
\newcommand{\wt}{\mathrm{wt}}
\newcommand{\VC}{Z_{\mathrm{vc}}}
\newcommand{\poly}{\operatorname{poly}}
\newcommand{\cutpoly}{\operatorname{cut}}
\newcommand{\hh}{\widehat h}
\newcommand{\hg}{\widehat g}
\newcommand{\Jhat}{\widehat J}
\newcommand{\Stil}{\widetilde S}
\newcommand{\calE}{\mathcal E}
\newcommand{\calL}{\mathcal L}
\newcommand{\calT}{\mathcal T}

\title{Maximum-Weight Matching in Random-Order Streams:\\
A One-Pass $(1-\epsilon)$-Approximation}
\author{Alireza Farhadi$^{1}$ \qquad Mohammadtaghi Hajiaghayi$^{2}$ \qquad Mohammad Mahdavi$^{2}$\\[3pt]
{\small $^{1}$Profound \qquad $^{2}$University of Maryland, College Park}}
\date{December 2026}
\begin{document}
\maketitle
\begin{abstract}
We study maximum-weight matching in the random-order streaming model,
where the edges of an $n$-vertex graph arrive one at a time in a
uniformly random order. We give a randomized one-pass approximation
scheme: for every fixed $\epsilon>0$, the algorithm returns a matching
of weight at least $(1-\epsilon)$ times the optimum with high
probability. For word-sized weights and inverse-polynomial failure
probability, it uses near-linear space in $n$ on general graphs. For
arbitrary positive rational weights, the space bound accounts for their
binary encoding. Neither the
stream length nor the range of the weights is needed in advance.

The algorithm uses a short prefix of the stream to decide which future
edges to keep, and thereby builds a sparse subgraph that contains a
near-optimal matching. The main difficulty is that each round of this
process needs information about edges that earlier rounds have selected
but that have yet to arrive. Our key technical tool is a sampling lemma
that retains a sparse graph containing a large matching on a set of
vertices that is defined only by local information in this partially
observed subgraph. On bipartite graphs we use the lemma to find many
disjoint edges that violate the constraints of the matching dual; on
general graphs a random bipartition exposes short augmenting paths and
cycles along which the current matching can be improved.
\end{abstract}

\section{Introduction}
Maximum matching is one of the central problems of graph streaming. In
the \emph{one-pass semi-streaming} model~\cite{fkmsz}, the edges of an
$n$-vertex graph arrive one at a time, and the algorithm must produce
its answer after a single pass using $n\cdot\poly\log n$ bits of
space. This is enough to store a matching, but far too little to store
a dense graph.

For adversarial edge order, greedily maintaining a maximal matching
gives a $1/2$-approximation for unweighted matching. Assadi, Jiang, and
Xiang~\cite{ajx} ruled out every fixed approximation ratio above
$1/2$ for randomized one-pass semi-streaming algorithms with constant
success probability, even on bipartite graphs with a perfect matching.
We consider the model in which the adversary chooses
the graph and its edges arrive in a \emph{uniformly random order}.
In this model every prefix of the stream is a uniform sample
of the graph, and a line of work
\cite{kmm,fhmrr,gkms,bernstein,ab,hw} has shown that one pass can
indeed do much better than $1/2$. For unweighted graphs, Bernstein
obtained a $(2/3-\epsilon)$-approximation~\cite{bernstein}, and Assadi
and Behnezhad improved the ratio to $2/3+\epsilon_0$ for an absolute
constant $\epsilon_0>0$~\cite{ab}. For weighted graphs, Hashemi and
Wrzos-Kaminska obtained a $(2/3-\epsilon)$-approximation~\cite{hw}.
In the first version of this paper
(\href{https://arxiv.org/abs/2611.17248v1}{arXiv:2611.17248v1}),
we used historical deletion bounds to obtain a deterministic
one-pass algorithm with approximation ratios of $0.71295$ for unweighted
matching and $0.71295-\epsilon$ for weighted matching.

\begin{samepage}
These results leave open the most basic question about the model:
\begin{quote}
\emph{Can a single pass over a random-order stream compute a
$(1-\epsilon)$-approximate maximum matching in semi-streaming space,
for every fixed $\epsilon>0$?}
\end{quote}
\end{samepage}
We answer this question affirmatively for general graphs with word-sized
weights. For arbitrary positive rational weights, the space bound below
accounts for their binary encoding.
Throughout, $\nu_w(G)$ denotes the maximum weight of a matching in $G$
and $\nu(G)$ its maximum cardinality.

\begin{theorem}[Main result]\label{thm:main}
Let $G$ be a simple graph on a known vertex set $[n]$, with $n\ge 2$
and positive rational edge weights. Its edges arrive in a uniformly
random order. For every fixed $0<\epsilon<1$ and every $0<\delta<1$,
a randomized one-pass algorithm returns a matching $M$
satisfying
\[
             \Pr\big[w(M)\ge(1-\epsilon)\nu_w(G)\big]\ge1-\delta.
\]
The probability is over the stream order and the algorithm's random choices.
The algorithm uses $O_\epsilon(n\Lambda^{C_\epsilon}(B+\Lambda))$
bits, where $\Lambda=1+\log(2n/\delta)$ and $B$ is the maximum width
of a numerator or denominator in the binary input ($B=1$ on the empty
input). The algorithm
requires neither the stream length nor bounds on the weights in advance.
Its running time is polynomial in $n,B,\Lambda$ for fixed $\epsilon$.
\end{theorem}

\subsection{Related work}\label{sec:related}
\paragraph{One pass, adversarial order.}
The semi-streaming model was introduced by Feigenbaum, Kannan,
McGregor, Suri, and Zhang~\cite{fkmsz}, who observed that a maximal
matching gives a $1/2$-approximation for unweighted matching. For
weighted matching, Paz and Schwartzman gave a
$(1/2-\epsilon)$-approximation~\cite{ps}. For unweighted bipartite graphs, earlier upper bounds
on the achievable ratio include $2/3$~\cite{gkk} and
$1/(1+\ln 2)$~\cite{kapralov}. Assadi, Jiang, and Xiang~\cite{ajx}
rule out every fixed $(1/2+\gamma)$-approximation, $\gamma>0$, with
constant success probability.

\paragraph{One pass, random order.}
Konrad, Magniez, and Mathieu were the first to beat $1/2$ in
random-order streams~\cite{kmm}. Farhadi, Hajiaghayi, Mai, Rao, and
Rossi improved the ratio substantially~\cite{fhmrr}, and Gamlath, Kale,
Mitrovi\'c, and Svensson beat $1/2$ for weighted
matching~\cite{gkms}. Bernstein obtained a $(2/3-\epsilon)$-approximation for
unweighted matching~\cite{bernstein}. Assadi and Behnezhad then broke the $2/3$
barrier with a $(2/3+\epsilon_0)$-approximation~\cite{ab}, and Hashemi
and Wrzos-Kaminska extended the $2/3-\epsilon$ guarantee to weighted
graphs~\cite{hw}. Our result closes this line of work for every fixed
$\epsilon$. The dependence of our space bound on $\epsilon$ cannot be
improved to $n\cdot\poly(\epsilon^{-1},\log n)$: for
$\epsilon_n=\Theta(1/\log n)$ chosen as in~\cite{ab}, even unweighted
bipartite matching requires $n^{1+\Omega(1/\log\log n)}$ space for a
$(1-\epsilon_n)$-approximation.

\paragraph{Multiple passes.}
A $(1-\epsilon)$-approximation was previously known only with more
than one pass, even in adversarial order: the number of passes depends
on $\epsilon$ but not on $n$~\cite{mcgregor,ag,fmu}.
Theorem~\ref{thm:main} shows that under random arrivals a single pass
suffices.

\paragraph{Tools.}
We build on three standard ingredients. Local algorithms that compute
near-optimal matchings on bounded-degree graphs go back to Nguyen and
Onak~\cite{no}. Short weighted augmentations are a classical tool for
approximate weighted matching~\cite{pettie-sanders}. The reduction
from arbitrary weights to bounded integer weights by weight classes is
due to Gupta and Peng~\cite{gp} and was used in random-order streams
by~\cite{hw}.

\subsection{Technical overview}\label{sec:overview}
\paragraph{Why a uniform sample is not enough.}
A random-order stream hands the algorithm a uniform sample of the
graph in every prefix, so the most naive strategy is to store a
uniform sample of $n\cdot\poly\log n$ edges. This can lose half of the
optimum. Take $K_{k,k}$ and attach a private leaf to each of its
$2k$ vertices. The leaf edges form a matching of size $2k$, but a
uniform sample of $O(k\log^c k)$ edges contains only $O(\log^c k)$ leaf
edges in expectation for fixed $c$. With $t$ leaf edges in the sample,
its matching number is at most $k+t/2$. The algorithm must therefore
decide \emph{which} edges are worth keeping, and it must do so before
most of them have arrived.

\paragraph{The framework: learn in a short prefix, collect in the suffix.}
For this overview, suppose that the stream length is known and that
the weights are integers in $[1,W]$ for a fixed $W$. The algorithm
maintains a sparse subgraph $H$ of the input, initially empty. It
never decides about an edge by looking at the edge in hindsight.
Instead, during a short \emph{training prefix} it fixes a sequence of
\emph{selection rules}, each of which assigns a retention probability
to every edge that arrives later. The subgraph $H$ is the set of
edges selected by the rules fixed so far. At any time the algorithm
stores the edges of $H$ that have already arrived, together with the
rules that will select the rest.

\begin{figure}[t]
\centering
\begin{tikzpicture}[x=1cm,y=1cm,font=\footnotesize,
  pv/.style={circle,draw=black!70,line width=0.6pt,fill=pfill,inner sep=0pt,minimum size=15pt},
  he/.style={line width=1.6pt,black!75},
  ve/.style={line width=1.5pt,viol,densely dashed},
  wl/.style={fill=white,inner sep=1.2pt,text=black}]
\definecolor{pfill}{HTML}{CDE7F7}
\definecolor{viol}{HTML}{D55E00}
\node[pv] (l1) at (0,3.2) {$2$};
\node[pv] (l2) at (0,1.9) {$1$};
\node[pv] (l3) at (0,0.6) {$0$};
\node[pv] (r1) at (4.2,3.8) {$1$};
\node[pv] (r2) at (4.2,2.5) {$2$};
\node[pv] (r3) at (4.2,1.2) {$0$};
\node[pv] (r4) at (4.2,-0.1) {$0$};
\draw[he] (l1) -- node[wl,pos=0.5] {$3$} (r1);
\draw[he] (l2) -- node[wl,pos=0.72] {$2$} (r1);
\draw[he] (l2) -- node[wl,pos=0.6] {$3$} (r2);
\draw[he] (l3) -- node[wl,pos=0.6] {$2$} (r2);
\draw[ve] (l1) -- node[wl,pos=0.9,text=viol] {$4$} (r3);
\draw[ve] (l3) -- node[wl,pos=0.5,text=viol] {$3$} (r4);
\node[anchor=east] at (-0.45,3.2) {$u_1$};
\node[anchor=east] at (-0.45,1.9) {$u_2$};
\node[anchor=east] at (-0.45,0.6) {$u_3$};
\node[anchor=west] at (4.65,3.8) {$v_1$};
\node[anchor=west] at (4.65,2.5) {$v_2$};
\node[anchor=west] at (4.65,1.2) {$v_3$};
\node[anchor=west] at (4.65,-0.1) {$v_4$};
\draw[he] (6.3,3.5) -- (7.1,3.5);
\node[anchor=west,align=left,text width=6.2cm] at (7.25,3.5)
  {edge of $H$: covered, $y_u+y_v\ge w_{uv}$};
\draw[ve] (6.3,2.6) -- (7.1,2.6);
\node[anchor=west,align=left,text width=6.2cm] at (7.25,2.6)
  {optimum edge outside $H$: violated};
\node[anchor=west,align=left,text width=7.2cm] at (6.3,1.2)
  {$u_1v_3$:\ \ deficit $4-(2+0)=2$\\[2pt]
   $u_3v_4$:\ \ deficit $3-(0+0)=3$\\[4pt]
   total price $=6=\nu_w(H)$};
\end{tikzpicture}
\caption{Dual prices show which edges are missing (Idea~1). The number
in a vertex is its price and the number on an edge is its weight. The
prices cover every edge of the current subgraph $H$ and their total
equals $\nu_w(H)$. Edges of an optimum matching that are not in $H$
can violate the dual constraints. Adding a matching of edges with
large deficit forces the potential up. We can test an edge using its
weight and its endpoint prices. For instance, $u_3v_4$ has weight $3$
and both endpoints have price $0$. This is
the form of the price-threshold tests in Idea~2.}
\label{fig:prices}
\end{figure}
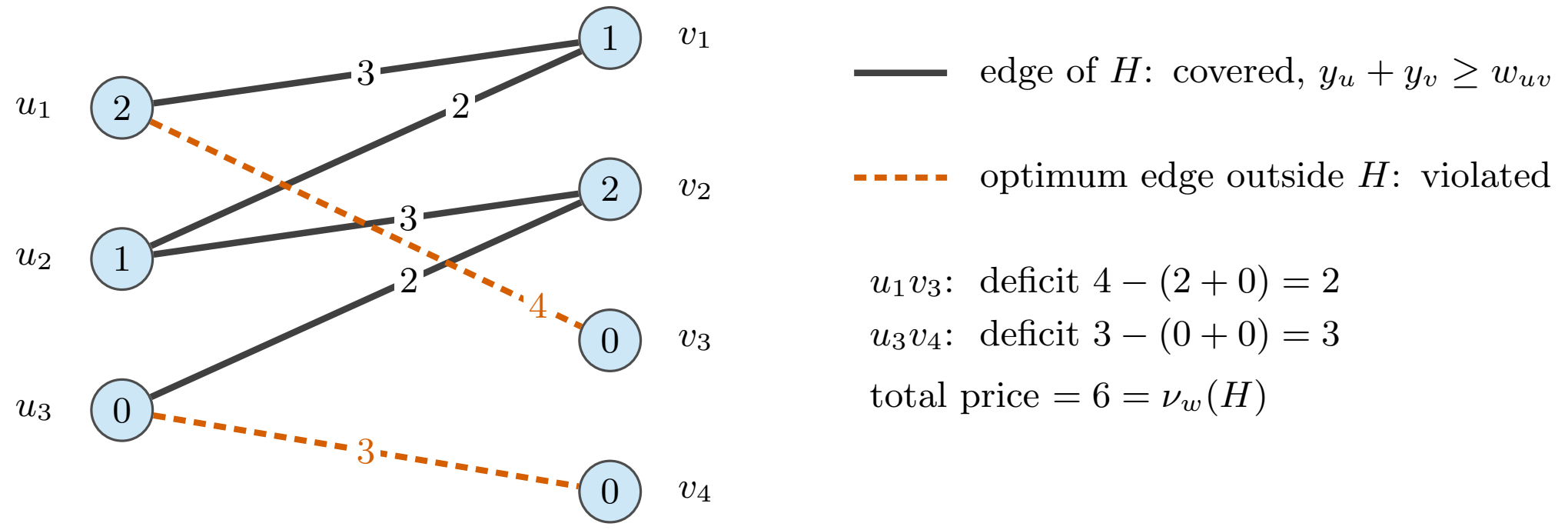

Each new rule is designed to add edges that increase the matching
weight of $H$. Since $\nu_w(H)\le\nu_w(G)$, a constant number of rules
with constant relative progress suffices. The catch is that a new rule
should depend on $H$, and most of $H$ lies in the future: only its
rules are known. The three ideas below show which edges to add, how
to add them without seeing $H$, and how to handle general graphs.
\paragraph{Idea 1: Which edges would help? Dual prices and a potential.}
Suppose first that the input is bipartite. By LP duality for bipartite
matching, $H$ admits vertex prices (a fractional weighted vertex
cover) of total cost $\nu_w(H)$ with $y_u+y_v\ge w_{uv}$ on every edge
of $H$. If $\nu_w(H)$ is far below the optimum, then many edges of an
optimum matching must violate these constraints. The natural plan is
to add violated edges to $H$. However, adding violated edges need not
immediately increase $\nu_w(H)$, since a different price vector may
cover them at the same cost (Figure~\ref{fig:prices} shows an
example of prices and violated edges). We therefore measure progress by a
smoothed version of the dual: we add a quadratic penalty on violations
and a small quadratic regularizer on the prices, and use the minimum of
this objective as a \emph{potential}. Adding a matching of edges with
large violations forces the potential up by an amount proportional to
the size of the matching. The potential is bounded by a constant times
the optimum, so a constant number of rounds suffices.

The smoothing has a second benefit. On a bounded-degree graph the
regularized objective is minimized, up to a small error, by a constant
number of projected gradient steps. Each vertex can therefore
approximate its price from a bounded-radius neighborhood: the price is
a \emph{local} function of the graph. Removing a small set of vertices
from the sparse subgraph $H$ bounds its degree at little cost in
matching weight.

\paragraph{Idea 2: How can we find those edges in one pass? Sampling
on a locally defined vertex set.}
We round the two endpoint prices to a fixed grid. Some pair of grid
thresholds identifies many disjoint edges whose weights exceed the sum
of the thresholds. It thus suffices to solve the following problem for
each pair of thresholds: \emph{keep a sparse set of future edges that
contains a large matching among the vertices whose price is below a
threshold}. The obstacle is that when a candidate edge arrives, its
endpoints' prices may still depend on unseen edges of $H$, so the
algorithm cannot simply test the endpoints.

Our sampling lemma (Lemma~\ref{lem:acquire}) overcomes this obstacle
using only an \emph{estimate} of how many eligible neighbors each
vertex has. Since eligibility is local, it can be written in terms of
counts of constant-size subgraphs of $H$, and such counts can be
estimated from a sample of $H$ that arrives during training. The
algorithm then processes degree scales from high to low. At each
scale it raises the sampling probability of all edges between
still-active vertices, and it deactivates a vertex once it has
accumulated nearly unit \emph{load} from its eligible neighbors: such a
vertex is expected to receive about one sampled edge to an eligible
neighbor, so its incident edges have received enough sampling
probability. Most edges of any fixed matching
on the eligible vertices either meet such a saturated vertex or remain
in a sparse graph that can be stored in full. Sampling from the
resulting edge probabilities preserves a constant fraction of the
matching. Sections~\ref{sec:sampling} and~\ref{sec:estimation} prove
the sampling lemma.

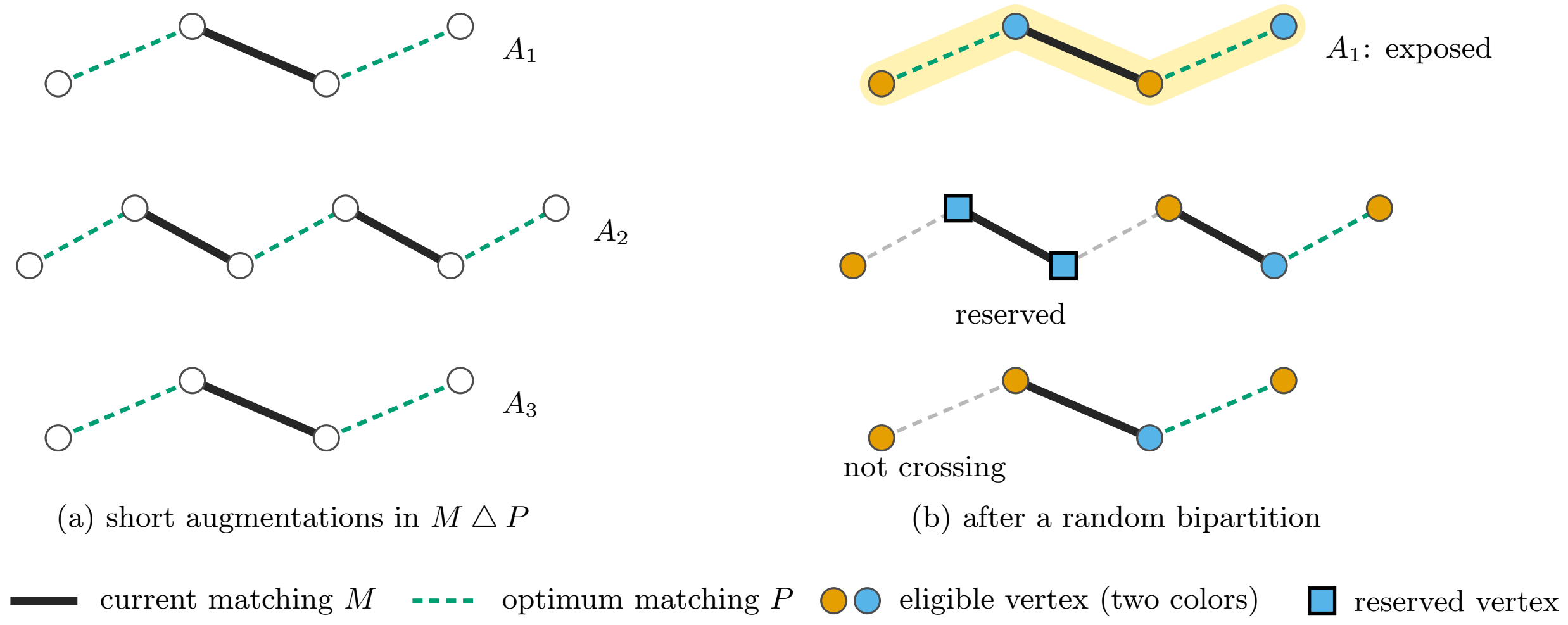
\begin{figure}[t]
\centering
\begin{tikzpicture}[x=1cm,y=1cm,font=\footnotesize,
  v/.style={circle,draw=black!70,line width=0.6pt,fill=white,inner sep=0pt,minimum size=7.5pt},
  vr/.style={v,fill=vred},
  vb/.style={v,fill=vblue},
  sr/.style={rectangle,draw=black,line width=1pt,fill=vred,inner sep=0pt,minimum size=7.5pt},
  sb/.style={rectangle,draw=black,line width=1pt,fill=vblue,inner sep=0pt,minimum size=7.5pt},
  m/.style={line width=2.4pt,black!85},
  opt/.style={line width=1.3pt,popt,densely dashed},
  off/.style={line width=1.1pt,black!28,densely dashed},
  hl/.style={line width=13pt,line cap=round,line join=round,hlcol}]
\definecolor{vred}{HTML}{E69F00}
\definecolor{vblue}{HTML}{56B4E9}
\definecolor{popt}{HTML}{009E73}
\definecolor{hlcol}{HTML}{FFF2B3}
\newcommand{\coords}{
  \coordinate (a1) at (0.3,3.4); \coordinate (a2) at (1.7,4.0);
  \coordinate (a3) at (3.1,3.4); \coordinate (a4) at (4.5,4.0);
  \coordinate (u1) at (0,1.5); \coordinate (u2) at (1.1,2.1); \coordinate (u3) at (2.2,1.5);
  \coordinate (u4) at (3.3,2.1); \coordinate (u5) at (4.4,1.5); \coordinate (u6) at (5.5,2.1);
  \coordinate (x1) at (0.3,-0.3); \coordinate (x2) at (1.7,0.3);
  \coordinate (x3) at (3.1,-0.3); \coordinate (x4) at (4.5,0.3);}
\begin{scope}
\coords
\draw[opt] (a1)--(a2); \draw[m] (a2)--(a3); \draw[opt] (a3)--(a4);
\draw[opt] (u1)--(u2); \draw[m] (u2)--(u3); \draw[opt] (u3)--(u4); \draw[m] (u4)--(u5); \draw[opt] (u5)--(u6);
\draw[opt] (x1)--(x2); \draw[m] (x2)--(x3); \draw[opt] (x3)--(x4);
\foreach \n in {a1,a2,a3,a4,u1,u2,u3,u4,u5,u6,x1,x2,x3,x4}{\node[v] at (\n) {};}
\node[anchor=west] at (4.8,3.75) {$A_1$};
\node[anchor=west] at (5.75,1.85) {$A_2$};
\node[anchor=west] at (4.8,0.05) {$A_3$};
\node at (2.75,-1.15) {(a) short augmentations in $M\mathbin\triangle P$};
\end{scope}
\begin{scope}[shift={(8.6,0)}]
\coords
\draw[hl] (a1)--(a2)--(a3)--(a4);
\draw[opt] (a1)--(a2); \draw[m] (a2)--(a3); \draw[opt] (a3)--(a4);
\draw[off] (u1)--(u2); \draw[m] (u2)--(u3); \draw[off] (u3)--(u4); \draw[m] (u4)--(u5); \draw[opt] (u5)--(u6);
\draw[off] (x1)--(x2); \draw[m] (x2)--(x3); \draw[opt] (x3)--(x4);
\node[vr] at (a1) {}; \node[vb] at (a2) {}; \node[vr] at (a3) {}; \node[vb] at (a4) {};
\node[vr] at (u1) {}; \node[sb] at (u2) {}; \node[sb] at (u3) {}; \node[vr] at (u4) {}; \node[vb] at (u5) {}; \node[vr] at (u6) {};
\node[vr] at (x1) {}; \node[vr] at (x2) {}; \node[vb] at (x3) {}; \node[vr] at (x4) {};
\node[anchor=west] at (4.8,3.75) {$A_1$: exposed};
\node[anchor=north,align=center] at (1.65,1.25) {reserved};
\node[anchor=north,align=center] at (0.75,-0.35) {not crossing};
\node at (2.75,-1.15) {(b) after a random bipartition};
\end{scope}
\begin{scope}[shift={(-0.2,-2.0)}]
\draw[m] (0,0)--(0.7,0); \node[anchor=west] at (0.8,0) {current matching $M$};
\draw[opt] (4.2,0)--(4.9,0); \node[anchor=west] at (5.0,0) {optimum matching $P$};
\node[vr] at (8.6,0) {}; \node[vb] at (8.95,0) {}; \node[anchor=west] at (9.15,0) {eligible vertex (two colors)};
\node[sb] at (13.7,0) {}; \node[anchor=west] at (13.9,0) {reserved vertex};
\end{scope}
\end{tikzpicture}
\caption{Why a random bipartition suffices (Idea~3). (a)~If the current
matching $M$ is far from optimal, $M\mathbin\triangle P$ contains many
disjoint short augmentations; three are shown. (b)~Each vertex receives
one of two colors. Only crossing edges are used. The monochromatic
edge of $M$ in $A_2$ is kept and its endpoints are reserved, which
hides $A_2$; one edge of $P$ in $A_3$ does not cross, which hides
$A_3$. All edges of $A_1$ cross, so $A_1$ survives: it uses only
eligible vertices and lies in a bipartite graph that is disjoint from
the reserved part of $M$. Each short augmentation survives with
constant probability, so a constant fraction of them does.}
\label{fig:cut}
\end{figure}

\paragraph{Idea 3: Why does a random bipartition suffice for general
graphs?}
In general graphs, vertex prices alone need not certify the optimum
matching weight. We reduce to the bipartite case by
coloring each vertex independently with one of two colors and keeping
only the edges that cross the bipartition. Used naively, this loses
half the weight of a fixed matching in expectation. We instead start
with a locally computed matching $M$ that is close to optimal in $H$
and use the bipartition only to \emph{improve} $M$
(Figure~\ref{fig:cut}). If $M$ is far from optimal in the input, its
symmetric difference with an optimum matching contains many disjoint
\emph{short augmentations}: alternating paths and cycles of constant
length whose flip increases the weight. Each of them lies entirely
across a random bipartition with constant probability. We keep the
edges of $M$ that do not cross the bipartition and reserve their
endpoints. The augmentations that cross entirely avoid the reserved
vertices, so all of them live in the remaining bipartite graph, where
the price argument applies. Since $M$ is computed by a local rule, the
set of non-reserved vertices is again locally defined, which is exactly
the setting of the sampling lemma. Section~\ref{sec:paper-reduction}
gives the iteration.

\paragraph{Removing the assumptions.}
Section~\ref{sec:implementation} removes the two simplifying
assumptions. For an unknown stream length, we run copies of the
algorithm on geometrically increasing guesses of the length. For
arbitrary weights, we partition the edges into geometric weight
classes, run the bounded-weight algorithm on blocks of consecutive
classes, and merge the resulting matchings greedily from the heaviest
block to the lightest.

\paragraph{The algorithm in one page.}
Algorithm~\ref{alg:overview} summarizes the bounded-weight algorithm.
The terms \emph{trimming}, \emph{local rule}, and \emph{selection rule}
are defined in Section~\ref{sec:prelim}.

\begin{algobox}[!tb]
\small
\textbf{Input:} a random-order stream of edges with integer weights in
$[1,W]$ and known length; accuracy $\epsilon$.\\
\textbf{Maintained:} a sparse subgraph $H$, given by the stored edges
and the selection rules fixed so far. Initially $H=\varnothing$.
\begin{enumerate}[leftmargin=1.6em,itemsep=2pt,topsep=3pt]
\item \textbf{Phases.} Repeat for a constant number of phases, all
inside the training prefix:
  \begin{enumerate}[leftmargin=1.4em,itemsep=1pt,topsep=1pt,label=(\alph*)]
  \item \emph{Trim.} Sample the edges of $H$, delete the endpoints of
  the sample, and delete the vertices whose remaining degree exceeds a
  cutoff. Call the result $H_0$.
  \item \emph{Local matching and random bipartition.} Fix the random
  bits of a local rule that computes a near-optimal matching $M$ of
  $H_0$, and color each vertex independently with one of two colors.
  Reserve the endpoints of the monochromatic edges of $M$; all other
  vertices are \emph{eligible}. These choices determine the eligible
  set $F$ even though some edges of $H$ have yet to arrive.
  \item \emph{Rounds.} Keep the bipartition and $F$ fixed for a
  constant number of rounds. In each round, trim the current subgraph
  afresh, reevaluate the eligibility rule on this trim, and compute
  local vertex prices. For each pair of price thresholds on a fixed
  grid, apply Algorithm~\ref{alg:sample} to the remaining crossing edges
  outside $H$ whose weight exceeds the sum of the thresholds by the
  prescribed margin. The local test accepts eligible vertices whose
  prices are at most their thresholds. Add the new rules to $H$.
  \end{enumerate}
\item \textbf{Every arrival.} Apply the rules already fixed, in the
order in which they were created. If some rule keeps the edge, store
one copy together with the identities of the rules that selected it.
\item \textbf{Output.} On the stored graph, start from the empty
matching and repeatedly flip a short augmentation until none remains.
\end{enumerate}
\caption{The bounded-weight algorithm (overview). A parallel buffer
handles inputs small enough to be stored in full.}
\label{alg:overview}
\end{algobox}

\subsection{Organization}
Section~\ref{sec:prelim} fixes notation and the basic conventions for
random-order streams. Section~\ref{sec:paper-reduction} states the
sampling lemma and uses it to prove the bounded-weight case
(Theorem~\ref{thm:bounded-general}). Section~\ref{sec:sampling}
describes the sampling rule and proves the sampling lemma, assuming two
estimation statements that are proved in Section~\ref{sec:estimation}.
Section~\ref{sec:implementation} handles arbitrary weights, unknown
stream length, and the space and time bounds, completing the proof of
Theorem~\ref{thm:main}.

\section{Preliminaries}\label{sec:prelim}

\paragraph{Graphs and matchings.}
All graphs are simple and have vertex set $V=[n]$. We identify a
subgraph with its edge set and write $|H|$ for the number of edges of
a subgraph $H$. For a vertex set $F$, $H[F]$ is the subgraph induced
on $F$, and $H-X$ is the graph obtained by deleting the vertices of
$X$. For a weight function $w$ and an edge set $M$, $w(M)$ is the
total weight of $M$. We write $\nu(H)$ for the maximum cardinality and
$\nu_w(H)$ for the maximum weight of a matching in $H$, and
abbreviate
\[
      r=\nu(G),\qquad \omega=\nu_w(G)
\]
for the input graph $G$. With integer weights in $[1,W]$ we have
$r\le\omega\le Wr$.

For matchings $M$ and $P$, each component of the symmetric difference
$M\mathbin\triangle P$ is a path or an even cycle whose edges alternate
between $M$ and $P$. \emph{Flipping} a set $A$ of edges in $M$ means
replacing $M$ by $M\mathbin\triangle A$. An \emph{augmentation} of $M$
is an alternating path or even cycle whose flip yields a matching of
larger weight; its \emph{gain} is the increase in weight.

\paragraph{Asymptotic notation and probability.}
Subscripts on $O$, $\Omega$ indicate parameters on which the hidden
constants may depend. Throughout, $\delta$ is the target failure
probability and $\Lambda=1+\log(2n/\delta)$. We use the Chernoff
bound for sums of independent indicators, tail bounds for binomial and
hypergeometric variables, and McDiarmid's bounded differences
inequality. Table~\ref{tab:notation} collects the notation that
recurs across sections.

\begin{table}[t]
\centering
\small
\begin{tabular}{@{}ll@{}}
\toprule
Symbol & Meaning \\
\midrule
$r,\ \omega$ & matching number $\nu(G)$ and maximum matching weight $\nu_w(G)$ of the input\\
$W$ & upper bound on the integer weights (Sections~\ref{sec:paper-reduction}--\ref{sec:estimation})\\
$\Lambda$ & $1+\log(2n/\delta)$\\
$H$, $J$ & reference graph (already selected, partly unseen) and candidate graph (unseen)\\
$p$, $d$ & sampling rate and degree cutoff used for trimming\\
$X$, $H'$, $H_0$ & endpoints of the $p$-sample of $H$; $H'=H-X$; $H_0$ is $H'$ after the degree cutoff\\
$F$ & eligible vertices: the set accepted by a local rule on $H_0$\\
$E_\times$ & edges crossing the current bipartition\\
$\Phi$ & potential: minimum of the regularized dual objective (Section~\ref{sec:eligible})\\
$D_i$, $a_i$ & degree scale and sampling increment of stage $i$ (Section~\ref{sec:sampling})\\
$T_i$, $T$ & active vertices at the start of stage $i$, and after the last stage\\
$h_v$ & surrogate for the indicator of $v\in F$, a combination of subgraph counts\\
$x_e$, $p_e$ & accumulated sampling mass and retention probability of a candidate edge $e$\\
\bottomrule
\end{tabular}
\caption{Notation used throughout the paper.}
\label{tab:notation}
\end{table}

\subsection{Random-order streams and arrival times}\label{sec:clocks}
It is convenient to analyze a random-order stream in continuous time.
Assign each edge an independent uniform \emph{arrival time} in $[0,1]$
and order the edges by arrival time; the resulting order is uniformly
random. The advantage of this view is a clean conditioning property:
conditional on the edges seen through a fixed time, all remaining
arrival times are independent and uniform on the remaining interval.

The algorithm sees only the permutation, not the times, but it can
simulate any fixed partition of $[0,1]$ into time intervals.
If $k$ edges remain and the next interval occupies a fraction $q$ of
the remaining time, the algorithm draws the size of the interval from
$\mathrm{Bin}(k,q)$ and reads that many edges. The block sizes have
the same multinomial law as under independent arrival times, and the
order conditional on the sizes is uniform in both models. Estimators
use the prescribed time fractions. An edge that is already stored can
join a fresh sample by an independent coin of the same bias.

\subsection{Selection rules}\label{sec:rules}
A \emph{selection rule} assigns to each edge that arrives after the
rule is fixed a retention probability; the edge is stored with this
probability, independently. Each stored edge records which rules
selected it.

For the analysis we may draw the independent coins of a rule for all
unseen edges at the moment the rule is fixed. This determines the
subgraph selected by the rule without revealing the remaining arrival
times. In particular, the edge set selected by earlier rules is a
fixed graph of which the algorithm has seen a part, and whose unseen
edges still have independent uniform arrival times on the remaining
interval.

\subsection{Trimming and local rules}\label{sec:trim}
The algorithm repeatedly needs to evaluate a function of a sparse
subgraph $H$ that it has not yet seen in full. We make this possible
by passing to a bounded-degree graph, on which the functions of
interest depend only on constant-radius neighborhoods.

\begin{definition}[Trimming]\label{def:trim}
Let $H$ be a graph, $p\in(0,1)$ a sampling rate, and $d$ a degree
cutoff. Sample each edge of $H$ independently with probability $p$ and
let $X$ be the set of endpoints of the sampled edges. Let $H'=H-X$,
and let $H_0$ be obtained from $H'$ by deleting every vertex whose
degree in $H'$ exceeds $d$. Deleted vertices are kept as isolated
vertices. We call $H_0$ the \emph{trimmed graph}.
\end{definition}

The initial sample serves two purposes. It removes most high-degree
vertices of $H$, because a vertex of degree $k$ survives with
probability $(1-p)^k$. It also makes the cutoff robust: after the
sample, vertices of high degree are so rare that the cutoff can be
tested approximately from a further sample of $H$
(Section~\ref{sec:estimation}).

\begin{definition}[Local rule and eligible set]\label{def:local}
A \emph{local rule of radius $R$} is a function that, for each vertex
$v$ of a labeled graph, outputs accept or reject depending only on the
radius-$R$ ball around $v$, including the labels (identifiers and
fixed random bits) of the vertices in the ball. This is the output of
an $R$-round algorithm in the LOCAL model of distributed computing.
We always require a local rule to accept every isolated vertex, and
to be evaluable on a constant-size neighborhood in polynomial time and
polylogarithmic space. The set $F$ of vertices accepted by a local
rule on the trimmed graph $H_0$ is called the \emph{eligible set}.
\end{definition}

\subsection{Inputs with a small matching number}\label{sec:small}
All concentration arguments below apply when $r$ exceeds a fixed
polynomial in $\Lambda=1+\log(2n/\delta)$. A parallel buffer covers
smaller $r$: the endpoints of a maximal matching cover every edge of
$G$, so $|E(G)|\le 2nr$, and an input with small $r$ can be stored in
full within the space bound.

\section{Matching with bounded integer weights}\label{sec:paper-reduction}
In this section and the next two, we assume that the stream length is
known and that the weights are integers in $[1,W]$ for a fixed $W$.
The goal is Theorem~\ref{thm:bounded-general}: one pass suffices to
store a sparse subgraph that contains a $(1-\epsilon)$-approximate
matching. The proof has three parts.
\begin{itemize}[leftmargin=*,itemsep=2pt,topsep=4pt]
\item Section~\ref{sec:acquire-statement} states the \emph{sampling
lemma}, the main technical tool. Its proof occupies
Sections~\ref{sec:sampling} and~\ref{sec:estimation}.
\item Section~\ref{sec:eligible} treats the bipartite problem
restricted to an eligible vertex set, using dual prices and a
potential (Lemma~\ref{lem:weighted-eligible-progress}).
\item Section~\ref{sec:cut} reduces general graphs to this bipartite
problem by a random bipartition, using the short augmentations of
Section~\ref{sec:short}.
\end{itemize}

\subsection{The sampling lemma}\label{sec:acquire-statement}
The sampling lemma addresses the following situation. A sparse
\emph{reference graph} $H$ has been selected by earlier rules; some of
its edges have arrived and are stored, and the rest are still to come.
A local rule on the trimmed graph $H_0$ defines an eligible set $F$,
which the algorithm cannot yet evaluate. A \emph{candidate graph} $J$
consists of unseen edges. We want to store a sparse part of $J$ that
preserves, up to a constant factor, the maximum matching of $J$ among
the eligible vertices.

\begin{lemma}[Sampling lemma: a matching on eligible vertices]\label{lem:acquire}
Let $H$ and $J$ be disjoint edge sets, each recognizable on arrival,
with $|H|\le Ar$ for a fixed $A$ and $\nu(H\cup J)\le r\le n$.
Suppose $r$ exceeds the threshold in Section~\ref{sec:small} and that
the following hold at the start of the call.
\begin{enumerate}[label=(\roman*),leftmargin=2.2em,itemsep=1pt,topsep=3pt]
\item Every edge of $H$ that has already arrived is in memory, and all
edges of $J$ are still unseen.
\item Conditional on the history and on the identities of the edges
in $H$ and $J$, the unseen arrival times are independent and uniform
on the remaining interval.
\end{enumerate}
Fix an accuracy $\eta>0$ and any positive upper bound on the fraction
of the stream that the call may read. Choose a sufficiently small
constant $p>0$ and any fixed degree cutoff $d$, trim $H$ with
parameters $p,d$ (Definition~\ref{def:trim}), and let $F$ be the
eligible set of a local rule fixed in advance
(Definition~\ref{def:local}).

There is an absolute constant $\beta>0$ with the following property.
Reading a prefix within the specified time budget suffices to fix a
selection rule that stores a subgraph $\Jhat\subseteq J$ from the
subsequent arrivals such that, with probability at least $1-\delta$,
\[
              |\Jhat|=O(r),\qquad
              \nu(\Jhat[F])\ge \beta\,\nu(J[F])-\eta r .
\]
\end{lemma}

\begin{remark}[Fine print for Lemma~\ref{lem:acquire}]\label{rem:acquire}
\begin{enumerate}[label=(\alph*),leftmargin=2.2em,itemsep=1pt,topsep=3pt]
\item Assumption~(ii) holds for the selection rules of
Section~\ref{sec:rules}, because their coins can be drawn when the
rules are fixed.
\item For word-sized labels the space is
$n\log^{O(1)}(2n/\delta)$.
\item The local rule may depend on $p$ and $d$.
\item A fixed finite family of tests, possibly with different
candidate graphs $J$, may share the same initial sample and degree
cutoff.
\item The hidden constants and the polylogarithmic exponents may
depend on all fixed parameters and rules.
\end{enumerate}
\end{remark}

\subsection{Short augmentations}\label{sec:short}
We need two standard facts, which we prove together. First, a matching
that is far from optimal admits many disjoint short augmentations.
Second, on a bounded-degree graph, a local rule finds a near-optimal
matching by repeatedly applying such augmentations. Local matching
algorithms on bounded-degree graphs go back to Nguyen and
Onak~\cite{no}; we use the following elementary weighted version.

\begin{lemma}[Short augmentations and a local matching]\label{lem:short-local-seed}
For each $0<a<1$ there are $L=L(a,W)$ and $c=c(a,W)>0$ with the
following properties.
\begin{enumerate}[label=(\arabic*),leftmargin=2.2em,itemsep=1pt,topsep=3pt]
\item Let $M$ and $P$ be matchings with $w(P)-w(M)\ge a\omega$. Then
$M\mathbin\triangle P$ contains at least $c\omega$ vertex-disjoint
alternating paths or even cycles of length at most $L$, such that
flipping any one of them in $M$ gives a matching whose weight is
larger by at least one.
\item On any subgraph $G'$ of maximum degree at most $d$, a
fixed-radius randomized local rule finds a matching of weight at
least $\nu_w(G')-a\omega$ except with probability
$\exp(-\Omega_{a,W,d}(r))$. The rule uses a constant number of
independent random bits per vertex.
\end{enumerate}
\end{lemma}

\begin{proof}
\emph{Part (1).}
Put $I=\lceil2W/a\rceil$. On each component of $M\mathbin\triangle P$ with more than $I$ edges
of $P$, delete every $I$th such edge in path or cyclic order. This costs at most
$W|P|/I\le a\omega/2$. Each
remaining component contains at most $2I$ edges of $P$, including
the one across the starting point of a cycle. Its flip is feasible
because only $P$-edges were cut. Hence components of positive gain
account for at least $a\omega/2$, at most $2IW$ apiece. We may take
$L=4I+1$ and $c=a/(4IW)$. Integrality makes each positive gain
at least one.

\emph{Part (2): the rule.}
On $G'$, start with the empty matching. In each round use fresh independent fair
bits to select vertices. In every component of
the selected induced subgraph with at most $L+1$ vertices and no
matching edge leaving it, replace the current edges by a maximum-weight
matching of the component. Break ties by vertex identifiers. This
rule is local and never decreases weight.

\emph{Part (2): the analysis.}
While the gap to $\nu_w(G')$ exceeds $a\omega$, take the disjoint
augmentations given by part~(1). The closed neighborhood of each
augmentation
meets those of only $O_{d,L}(1)$ others, since each such augmentation
contains a distinct vertex within distance two of its vertex set.
Greedily keep a constant fraction with disjoint closed neighborhoods.
For each retained augmentation, with probability at least
$2^{-(d+1)(L+1)}$ exactly its vertices are selected in its closed
neighborhood. It is then processed as a component and gains at least
one. These events are independent. A Chernoff bound gives
$\Omega_{a,W,d}(\omega)$ gains, except with the claimed probability.
Since every matching weighs at most $\omega$, a fixed number of
rounds suffices. Each round looks only a bounded distance to detect
a small component and its incident matching edges.
\end{proof}

\subsection{Weighted bipartite matching on eligible vertices}\label{sec:eligible}

\paragraph{Setting.}
At the start of a phase, let $H$ be the maintained subgraph, with
$|H|\le Ar$. Trim $H$ as in Definition~\ref{def:trim}: sample its
edges at rate $p$, delete their endpoints $X$ to obtain $H'$, and
delete the vertices of $H'$ of degree above $d$ to obtain $H_0$.
Trimming is cheap: for any fixed $a>0$, taking $p$ small and $d$ large
gives
\[
        \nu_w(H_0)\ge\nu_w(H)-a\omega
\]
with high probability. Indeed, each deleted vertex costs at most $W$,
the number of sampled endpoints is at most twice a binomial count of
mean $p|H|$, and at most $2Ar/d$ vertices exceed the degree cutoff.

Fix a bipartition of $V$ with crossing edges $E_\times$, and an
eligible set $F$ defined by a radius-$R$ local rule on $H_0$. As the
maintained graph $K\supseteq H$ grows, our target subgraph is
\[
            S(K)=K[F]\cap E_\times,
\]
the crossing edges of $K$ between eligible vertices. The next lemma
finds a matching in $S(K)$ that is nearly as heavy as any fixed
\emph{benchmark matching} $Q$ on $F$, even though $F$ can be evaluated
exactly only after all of $H$ has been seen.

\begin{lemma}[Progress on eligible vertices]\label{lem:weighted-eligible-progress}
Fix $0<\rho<1/4$ and a benchmark matching $Q$ on $F$ crossing the
bipartition, chosen without revealing future arrival times.
Starting from $|K|\le A'r$, a fixed number of sampling rounds produces
$K'\supseteq K$ with $|K'|=O(r)$. With failure probability at most $\delta$ (including the
initial sample of $H$),
\[
          \nu_w(S(K'))\ge w(Q)-\rho\omega
\]
whenever the already
arrived edges of $Q$ outside the maintained subgraph weigh at most
$\rho\omega/32$ at every call. The algorithm does not use $Q$. The
number of rounds depends only on $\rho$ and $W$.
\end{lemma}

\paragraph{Proof outline.}
We define a potential $\Phi$ on bipartite graphs, a smoothed version
of the matching dual, and show that each round in which $S(K)$ is
still deficient increases $\Phi(S(K))$ by $\Omega(\omega)$. The
proof has five steps: (1)~basic properties of $\Phi$ and of its
minimizing prices; (2)~adding a matching of edges with large dual
violation increases $\Phi$; (3)~the eligible set $F$ can be
re-evaluated on a fresh trim of the current graph up to a small error
set $Y$; (4)~a deficient $S(K)$ leaves many benchmark edges with large
violation; and (5)~a constant number of price-threshold tests, each
solved by the sampling lemma, captures a constant fraction of them.

\begin{proof}
Write $\tau=\rho/32$, $\lambda=\rho/8$, $m=\lambda/W$, and $k=1/\tau$.

\paragraph{Step 1: the potential and local prices.}
On a bipartite graph $\Gamma$ use the potential
\[
 \Phi(\Gamma)=\min_{0\le y\le W} f_\Gamma(y),\qquad
 f_\Gamma(y)=\sum_v\left(y_v+\tfrac m2 y_v^2\right)
        +\tfrac k2\sum_{uv\in \Gamma}[w_{uv}-y_u-y_v]_+^2,
 \qquad [x]_+=\max\{x,0\}.
\]
We call $y$ a vector of \emph{prices} and $[w_{uv}-y_u-y_v]_+$ the
\emph{deficit} of $uv$ at $y$: the amount by which the edge violates
the dual constraint $y_u+y_v\ge w_{uv}$. Thus $f_\Gamma$ is the dual
objective $\sum_v y_v$ plus a regularizer with parameter $m$ and a
penalty on deficits with parameter $k$. We use three properties.
\begin{itemize}[leftmargin=*,itemsep=2pt,topsep=4pt]
\item \emph{Upper bound.} By weighted bipartite matching duality there
is a feasible cover of total price $\nu_w(\Gamma)$, with prices clipped
at $W$. Thus $\Phi(\Gamma)\le (1+\lambda/2)\nu_w(\Gamma)$.
\item \emph{Monotonicity.} Adding edges can only increase $\Phi$.
\item \emph{Stability.} Two graphs agreeing outside $Y\subseteq V$
have potentials differing by at most $(1+\lambda/2)W|Y|$, by raising
every price in $Y$ to $W$.
\end{itemize}

On a graph of maximum degree at most $\Delta$, a local rule computes
prices $z$ within $\tau$ in supremum norm of the minimizer $y^*$.
Indeed, start at zero and iterate the projected gradient update
\[
 y'_v=\Pi_{[0,W]}\!\left(y_v-\sigma\left(1+my_v-
                k\sum_{uv\in \Gamma}[w_{uv}-y_u-y_v]_+\right)\right),
 \qquad \sigma=(m+k\max\{1,\Delta\})^{-1}.
\]
With $j$ positive incident deficits the unprojected update has
diagonal derivative $1-\sigma(m+kj)\ge0$ and off-diagonal derivatives
of total absolute value $\sigma kj$. By continuity across its linear
pieces and projection it contracts the supremum norm by $1-\sigma m$.
The minimizer is a fixed point, so a constant number of local steps
suffices for fixed $\Delta,\rho,W$. Isolates retain price zero.

Finally, at $y^*$ every edge of $\Gamma$ has deficit at most
$(1+\lambda)/k$. Indeed an endpoint priced $W$ has no deficit, and at
any lower price coordinate optimality gives
$k\sum_{uv\in \Gamma}[w_{uv}-y_u^*-y_v^*]_+\le 1+my_v^*$.

\paragraph{Step 2: violated matchings increase the potential.}
If $N$ is a matching outside $\Gamma$ with deficit at least $2\tau$ per
edge at $y^*$, then
\begin{equation}
      \Phi(\Gamma\cup N)-\Phi(\Gamma)\ \ge\ (m\tau^2/4)|N|.
      \label{eq:reduction-price-progress}
\end{equation}
Indeed $f_\Gamma(y)\ge\Phi(\Gamma)+(m/2)\|y-y^*\|_2^2$ by strong
convexity. For each $uv\in N$,
changing its two coordinates by at least $\tau$ in total absolute
value costs $m\tau^2/4$ in this quadratic term. Otherwise its new
edge penalty is at least $k\tau^2/2$. Disjointness lets us sum these
bounds at any $y$, including the new minimizer.

\paragraph{Step 3: evaluating the old eligibility rule on a fresh trim.}
One round of Lemma~\ref{lem:acquire} samples and trims the current
graph $K$, giving a bounded-degree graph $K_0=K-Z$. As for $H_0$, we
may make $|Z|\le\xi r$ with high probability for any fixed $\xi>0$.
The eligible set $F$ was defined on the old trimmed graph $H_0$, which
differs from the part of $H$ inside $K_0$. We therefore re-evaluate
the rule: on the $H$-edges of $K_0$, delete the original $X$, reapply
the original degree cutoff $d$, then evaluate the original local rule
with the same vertex labels. Let $F'$ be the result and
\[
           Y=Z\cup(F\mathbin\triangle F')
\]
the set of vertices where the two graphs or the two answers may
differ. We claim that $Y$ is small.

Removing $Z$ from $H'$ changes degrees only at $Z$ and its neighbors.
The old degree cutoff can thus change incident edges only within
distance two of $Z$, even when a vertex now falls below the cutoff.
The radius-$R$ answer can change only within distance $R+2$ of $Z$
in $H'$.
Write $\Ball_s^{H'}(v)$ for the closed radius-$s$ ball around $v$ in
$H'$. Apply Lemma~\ref{lem:est-censor} at the start of the phase,
before sampling $H$. On its success event, with $s=R+2$,
\[
 \sum_v (|\Ball_s^{H'}(v)|-1)^2\le C_{A,p,s}r,\qquad
 |Y|\le\sum_{v\in Z}|\Ball_s^{H'}(v)|
       \le (\xi+\sqrt{C_{A,p,s}\xi})r ,
\]
where the last step is the Cauchy--Schwarz inequality.
This holds for every $Z$, including one chosen using later observations.
Hence $|Y|\le t r$ for any required fixed $t>0$ by choosing $\xi$
sufficiently small.

We use the graph
\[
        \Stil=K_0[F']\cap E_\times
\]
as a proxy for $S(K)$: the two agree outside $Y$. All prices $z$ on
$\Stil$ are local functions on the same bounded-degree graph $K_0$, by
composition of the rules. Let $y^*$ be the minimizer of $f_{\Stil}$.

\paragraph{Step 4: a deficient graph leaves many edges with a large deficit.}
Suppose $\nu_w(S(K))<w(Q)-\rho\omega$. Among the edges of $Q$ still
available, discard those touching $Y$, costing at most $Wt\omega$.
Altogether the remaining matching weighs at least
$w(Q)-(\tau+Wt)\omega$. Its total price at $y^*$ is at most
\[
 \sum_v y_v^*\le\Phi(\Stil)\le\nu_w(\Stil)+(\lambda/2)\omega
             \le w(Q)-(\rho-\lambda/2-Wt)\omega .
\]
Take $Wt\le\tau$. The total deficit on this matching is then at
least $3\rho\omega/4$. Deficits below $8\tau=\rho/4$ contribute at most
$\rho\omega/4$ since $|Q|\le\omega$. As no deficit exceeds $W$, at
least $\rho\omega/(2W)$ edges have deficit at least $8\tau$. All lie
outside $K$, since any such edge of $K$ outside $Y$ would be in
$\Stil$, where every deficit is at most $(1+\lambda)/k<8\tau$.

\paragraph{Step 5: price-threshold tests and the sampling lemma.}
Take the grid of step $\tau$ on $[0,W]$, including $W$. For each
pair of grid values $(\pi_L,\pi_R)$ define the following
\emph{test}.
\begin{itemize}[leftmargin=*,itemsep=2pt,topsep=4pt]
\item A left endpoint is eligible for the test if it belongs to $F'$
and has $z$-price at most $\pi_L$; a right endpoint is eligible if it
belongs to $F'$ and has $z$-price at most $\pi_R$.
\item The candidate graph consists of the crossing edges outside $K$
of weight at least $\pi_L+\pi_R+4\tau$.
\end{itemize}
These eligibility rules are local and accept all isolates. Every edge
counted in Step~4 passes the test obtained by rounding its two
$z$-prices up to the grid, since each rounded price is at most its
true price plus $2\tau$. Conversely every admitted edge has true
deficit at least $2\tau$. Let $\calT$ be the family of tests; its
size depends only on $\rho$ and $W$. Thus one of the tests has a
benchmark matching with at least $\mu\omega$ edges, where
$\mu=\rho/(2W|\calT|)$.

Apply Lemma~\ref{lem:acquire} to all tests with accuracy $\beta\mu/4$.
Its guarantee, followed by deleting at most $tr$ matching edges
touching $Y$, gives a matching $N$ among the stored candidate edges
with $|N|\ge\beta\mu\omega/2$ in the successful test
when $t\le\beta\mu/4$. Each edge of $N$ has both endpoints in $F$
and deficit at least $2\tau$ at $y^*$. Add all the new sampling rules
to obtain $K^+$. Inequality~\eqref{eq:reduction-price-progress}
gives $\Phi(\Stil\cup N)-\Phi(\Stil)\ge g\omega$, where
$g=m\tau^2\beta\mu/8>0$. Transferring both potentials across $Y$ costs
at most $2(1+\lambda/2)Wtr$ by stability. Choose $t$ small enough that
this is at most $g\omega/2$. Since $S(K^+)\supseteq S(K)\cup N$,
monotonicity gives
\[
                 \Phi(S(K^+))\ge\Phi(S(K))+g\omega/2 .
\]

\paragraph{Order of parameters and the number of rounds.}
The gain $g$ depends only on $\rho,W$. Choose $t$ accordingly, then
$\xi$ from the original neighborhood bound, then the fresh sampling
rate and degree cutoff. Only now is the radius of the new price rule
needed. It affects the cost of a sampling call and the next graph's
size constant, neither of which enters $g$. Since the potential is
at most $(1+\lambda/2)\omega$ and $S(K)$ grows, a predetermined
$1+\lceil2(1+\lambda/2)/g\rceil$ rounds suffice. Each round adds
$O(r)$ edges. Sum the conditional failure probabilities over calls. Intersect the
result with the assumed bound on missed benchmark edges only afterwards.
\end{proof}

\subsection{General graphs via a random bipartition}\label{sec:cut}
\begin{theorem}[Bounded integer weights and known stream length]
\label{thm:bounded-general}
For fixed $W$ and $\epsilon>0$, a single random-order pass on a
general graph with integer weights in $[1,W]$ produces a subgraph
containing a matching of weight at least $(1-\epsilon)\nu_w(G)$
with probability at least $1-\delta$. The algorithm uses
$O_{W,\epsilon}(n\,\mathrm{polylog}(n/\delta))$ space. For large $r$
the subgraph has $O(r)$ edges and the sampling rules may be fixed
within any prescribed initial constant fraction of the stream.
Constants may depend on that fraction.
\end{theorem}

\begin{proof}
Assume $0<\epsilon<1/4$. Fix a maximum-weight matching $P$ before
the stream. Let $c,L$ be from Lemma~\ref{lem:short-local-seed} at
tolerance $\epsilon$ and set $\kappa=c2^{-L}/2$, $a=\rho=\kappa/8$.

\paragraph{The training prefix.}
Schedule all training before a fixed time $t_\star$ sufficiently small that
the weight of $P$ arriving before $t_\star$ is at most $\rho\omega/32$ except with the
assigned failure probability. Indeed its edge count is binomial with
parameters $|P|,t_\star$, each such edge weighs at most $W$, and $|P|\le
\omega$. Taking a further constant margin in $t_\star$ also bounds the number of edges in the initial intervals by a binomial tail.

\paragraph{One phase.}
Begin with $H=\varnothing$. In each phase trim $H$ as in
Section~\ref{sec:eligible}, then apply the local rule of
Lemma~\ref{lem:short-local-seed}
to $H_0$ at tolerance $a$. Its matching $M$ weighs at least
$\nu_w(H)-2a\omega$. Give each vertex an independent fair color.
Reserve both endpoints of every monochromatic edge of $M$ and let
$F$ be the other vertices. This is a local rule on $H_0$ accepting
isolates. Any crossing matching on $F$ is disjoint from the reserved
matching $M\setminus M_\times$, where $M_\times$ is the crossing
part of $M$.

\paragraph{The benchmark matching.}
When $\nu_w(H)<(1-\epsilon)\omega$, we have $w(P)-w(M)>\epsilon\omega$.
Lemma~\ref{lem:short-local-seed} gives at least $c\omega$ disjoint
short augmentations. Conditional on the
unordered graph, the prior rules and acceptance bits, the initial
sample of $H$ and the local matching bits, these augmentations can be
fixed before drawing the colors. Each crosses in its entirety with
probability at least $2^{-L}$, independently on the disjoint vertex
sets. By a Chernoff bound at least $\kappa\omega$ cross except with
probability $\exp(-\Omega_{\epsilon,W}(r))$. Their vertices belong
to $F$, since an augmentation contains each incident edge of $M$.
Flipping them in $M_\times$ gives a matching $Q$ on $F$ of weight
at least $w(M_\times)+\kappa\omega$. Every edge of $Q$ outside $H$
belongs to $P$. Hence the single initial bound on the weight of
already arrived edges of $P$ applies at every subsequent call. This comparison
uses no remaining arrival times.

\paragraph{Progress per phase.}
Apply Lemma~\ref{lem:weighted-eligible-progress} starting from
$K=H$, keeping $F$ and the bipartition fixed throughout its rounds.
Its resulting subgraph contains a crossing matching of weight at
least $w(Q)-\rho\omega$ together with all the reserved edges. Their
union weighs at least $\nu_w(H)+(\kappa-2a-\rho)\omega$. Use the
entire enlarged subgraph as the next $H$. Each deficient
phase thus raises the actual matching optimum by at least
$\kappa\omega/2$. The subgraphs grow, and their optima never exceed
$\omega$, so $1+\lceil 2/\kappa\rceil$ phases suffice. Their number
is fixed before choosing any subgraph size constants. Finally take
a union bound over the calls, or use the full-input buffer when $r$
is small.
\end{proof}

\begin{remark}[Bipartite graphs and unit weights]
For a supplied bipartition, Lemma~\ref{lem:weighted-eligible-progress}
applies directly with $H=\varnothing$, $F=V$, $Q=P$; no random
bipartition or phases are needed. Unit weights are the special case
$W=1$ in both arguments.
\end{remark}

\section{Sampling on a locally defined vertex set}\label{sec:sampling}
This section proves the sampling lemma (Lemma~\ref{lem:acquire}). The
goal is to retain edges on the eligible set $F$ before we know which
vertices belong to $F$.

\paragraph{The idea.}
Suppose for a moment that we knew $F$. Imagine raising the sampling
probability of all candidate edges gradually and at a common rate, and
freezing the edges at a vertex as soon as the vertex expects one
sampled edge to an eligible neighbor. A frozen vertex is
\emph{saturated}: its edges to eligible neighbors carry total
probability about one, which is what a matching needs, and freezing
keeps the expected number of stored edges per vertex bounded. We
implement a robust version of this process that needs only
\emph{estimates} of the number of eligible neighbors. The algorithm
sweeps over degree scales $D_0>D_1>\cdots$ from order $n$ down to a
constant. At scale $D_i$ it adds an increment $a_i\propto 1/D_i$ to
the sampling probability of every candidate edge whose endpoints are
both still \emph{active}. A vertex becomes inactive once its
\emph{load}, the total increment received from its eligible neighbors,
reaches one. So a vertex is deactivated at a scale proportional to its
number of eligible neighbors, and the stage at which a vertex leaves
the active set determines the sampling rate of its later incident
edges.

Since eligibility is not known, the load is estimated. A local rule
on the trimmed graph $H_0$ can be approximated by a nonnegative
\emph{score} $h_v$, which is a weighted sum of counts of constant-size
rooted subgraphs of $H'$. Subgraph counts have unbiased estimators
from an edge sample, and a fresh sample of $H'$ is available in every
time interval.

\subsection{The sampling rule}\label{sec:rule}
Fix one test and choose an internal tolerance $\gamma$ sufficiently
small compared with $\eta$. After the initial sample of $H$, recall
that $H'$ is the graph left by deleting the sample's endpoints and
$H_0$ is the trimmed graph obtained after the degree cutoff.
Proposition~\ref{prop:marker-estimation} approximates the indicator
$\mathbf1_{v\in F}$ on $H_0$ by a nonnegative score $h_v$ on $H'$. The
score is a weighted sum of counts of constant-size rooted subgraphs.
Sampling edges of $H'$ therefore gives an unbiased estimate of $h_v$.

\paragraph{Scales.}
Start with the active set $T_0=V$. For fixed $0<\theta\le\gamma$,
choose a large constant $A_0$ and set
\[
 D_0=2A_0n,\qquad D_{i+1}=D_i/(1+\theta),\qquad a_i=\theta A_0/D_i .
\]
At stage $i$, let $N_i(v)$ be the unseen candidate neighbors of $v$ in
$T_i$, taking $N_i(v)=\varnothing$ if $v\notin T_i$. Write
\[
       \deg_i(v)=|N_i(v)|,\qquad g_i(v)=\sum_{w\in N_i(v)}h_w
\]
for the stage-start degree of $v$ and its total neighbor score.
Algorithm~\ref{alg:sample} gives the rule; each stage uses three
successive time intervals.

\begin{algobox}[!tb]
\small
\textbf{Training (stages $i=0,1,\dots,s-1$).} Stage $i$ reads three
successive time intervals.
\begin{enumerate}[leftmargin=2.4em,itemsep=2pt,topsep=3pt,label=(\roman*)]
\item \emph{Estimate eligibility.} Sample edges of $H'$ at rate
$q_i=\Theta(D_i^{-\phi})$ for a small constant $\phi>0$. Use the first
interval for unseen edges and fresh coins for stored edges, as in the
initial sample. Multiply the coefficient of each sampled rooted
subgraph with $k$ edges by $q_i^{-k}$ to estimate $h_w$ at its root
$w$. Call this estimate $\hh_i(w)$.
\item \emph{Estimate loads.} In the second interval, sum
$\hh_i(w)/q_i$ over the sampled neighbors $w\in N_i(v)$. Call the
result $\hg_i(v)$. Remove $v$ from the active set when its cumulative
estimated load $\sum_{j\le i}a_j\,\hg_j(v)$ reaches one.
\item \emph{Degree test.} Among the remaining active vertices, count
the candidate edges arriving in a third interval of relative length
$q'_i=\Theta(q_i)$ measured from its own start. Remove vertices with
sampled degree above $q'_iD_i/2$ to obtain $T_{i+1}$.
\end{enumerate}
Stop at the first $s$ with $D_s\le D_*$ and write $T=T_s$.

\medskip
\textbf{Selection (every later candidate edge $e=uv$).} Set
\[
 x_e=\sum_{i:\,u,v\in T_i}a_i,\qquad
 p_e=\begin{cases}1 & u,v\in T,\\ \min\{1,x_e\}&\text{otherwise} ,\end{cases}
\]
and store $e$ with independent probability $p_e$.
\caption{The sampling rule for one test, with reference graph $H$ and
candidate graph $J$.}
\label{alg:sample}
\end{algobox}

Four comments on Algorithm~\ref{alg:sample} are in order.
\begin{itemize}[leftmargin=*,itemsep=2pt,topsep=4pt]
\item \emph{Unbiasedness.} The first two intervals of a stage have
equal length measured from the start of the stage. Reference and
candidate edges are disjoint, so $\hg_i(v)$ is an unbiased estimate of
$g_i(v)=\sum_{w\in N_i(v)}h_w$.
\item \emph{Time budget.} The stage intervals have total length
$O_{\phi,\theta}(D_*^{-\phi})$, so a sufficiently large constant $D_*$
fits any fixed budget after the initial sample.
\item \emph{Several tests.} A finite family of tests can use common
parameters and share the initial sample.
\item \emph{Memory.} The last active stage of each endpoint suffices to
compute $p_e$, so the rule is described by one array per vertex.
\end{itemize}

\subsection{Proof of the sampling lemma}
The proof uses two estimation statements,
Propositions~\ref{prop:marker-estimation}
and~\ref{prop:stage-estimation}, which are proved in
Section~\ref{sec:estimation}. Informally, the first says that the
score $h_v$ agrees with eligibility outside $O(\gamma r)$ vertices,
and the second says that the estimated loads are accurate outside an
exceptional set carrying $O(\gamma r)$ load. We use these estimates to
bound the number of stored edges by $O(r)$, track the load that
survives training, and show that the sampled graph contains a matching
covering a constant fraction of the vertices with surviving load near one.

\begin{proof}[Proof of Lemma~\ref{lem:acquire}]
Fix one test. All probability bounds are conditional on the history
before the call and the reference and candidate edge sets. For a finite
family of tests, divide the failure probability among them.

\paragraph{Notation.}
Set
$C=\{v\in F:h_v\ge 1/2\}$, $c_v=\mathbf1_{v\in C}$ and $U=V\setminus C$.
We think of $C$ as the \emph{confirmed eligible} vertices: those that
are eligible and whose score reflects it.
These sets depend only on the initial sample of reference edges. By
Proposition~\ref{prop:marker-estimation}, $|F\setminus C|=O(\gamma r)$ and
$|U|=O_A(r)$. Define the \emph{true load} and the \emph{eligible load}
of $v$ by
\[
   L_v=\sum_i a_i\,g_i(v)=\sum_i a_i\sum_{u\in N_i(v)}h_u,\qquad
   L_v^c=\sum_i a_i|N_i(v)\cap C| .
\]
Proposition~\ref{prop:stage-estimation} shows
that $L_v^c=O(1)$ outside a set of
exceptional vertices on which its sum is $O(\gamma r)$. Fix a vertex
cover $\VC$ of $H\cup J$ of size at most $2r$ by taking the endpoints
of a maximal matching of $H\cup J$.

\paragraph{Part 1: the number of stored edges.}
We bound the total mass $\sum_e x_e$ over the candidate edges of the
suffix by splitting them into three types.

\emph{Edges within $C$ and between $C$ and $U$.}
The total $x$-mass on suffix edges within $C$ is at
most $\sum_{v\in \VC\cap C}L_v^c=O(r)$, and that on edges between $C$
and $U$ is at most $\sum_{v\in U}L_v^c=O_A(r)$.

\emph{Edges inside $U$.} Charge each edge to the first of its endpoints to
leave the active set, breaking ties arbitrarily and treating the end of training
as a departure for surviving vertices. For a vertex $v\in U$ assigned edges at
scale $D_i$, each assigned edge has mass at most
$\sum_{k\le i}a_k\le(1+\theta)A_0/D_i$. The number of
these edges is at most the stage-start degree $\deg_i(v)$.
Vertices with $\deg_i(v)\le D_i$ are therefore charged $O(A_0)$ once.
For the rest, the estimate
$\sum_{v:\deg_i(v)>D_i}\deg_i(v)\le rD_i^{-3}$ of
Proposition~\ref{prop:stage-estimation} bounds the total charge
over all scales by $O(A_0r\sum_i D_i^{-4})$.

Thus
$\sum_e x_e=O(r)$. The remaining graph on $T$ has $O(rD_*)$ edges
by the final degree estimate. Conditional on training and the unordered
suffix, a Chernoff bound now gives $|\Jhat|=O(r)$ except with probability
$\exp(-\Omega(r))$.

\paragraph{Part 2: how much eligible load survives training?}
The loads $L_v^c$ count all candidate edges that are unseen at the
start of each stage, but some of these arrive during training and are
lost. We show that the loss is small. For $v\in C$, let $\ell_v$ be
the total $x$-mass on suffix edges
from $v$ into $C$. At the start of stage $i$ let $R_i$ be the remaining edges on
$T_i\cap C$. In the arrival-time model, conditional on the
history up to that start and the initial sample of reference edges, each edge
of $R_i$ has probability at most $2\alpha$ of arriving before the
fixed end of training, where $\alpha<1/8$ is the total training
fraction. These arrival times are independent at that start. A binomial tail bound and a union bound give at most
$4\alpha|R_i|+O(\Lambda)$ such arrivals at every stage.
Sum with coefficients $2a_i$. Since $\sum_i a_i|R_i|\le
\sum_{v\in \VC\cap C}L_v^c=O(r)$ and $\sum_i a_i=O(A_0/D_*)$,
choosing $\alpha$ small enough gives
\[
           \sum_{v\in C}(L_v^c-\ell_v)=O(\gamma r).
\]
Each difference is nonnegative. This bound conditions only on the history
at the start of the current stage.

Next we compare with the true loads.
Proposition~\ref{prop:stage-estimation} gives
$\sum_{v\in C}|L_v^c-L_v|=O(\gamma r)$. Outside the exceptional set,
$L_v\le1+O(\gamma)$, and $L_v\ge1-O(\gamma)$ at every removal due to
high estimated load.

\paragraph{Part 3: a maximum matching on eligible vertices.}
Fix a maximum matching $M_J$ of $J[F]$ conditional
on the initial sample of reference edges, before exposing any candidate arrival times.
It loses $O(\gamma r)$ edges to the training prefix by concentration.
Discard also its edges touching $F\setminus C$, an exceptional vertex,
or a vertex of $C$ removed by the degree test. These
cost $O(\gamma r)$ by the estimation propositions. Every remaining
edge is of one of two types:
\begin{itemize}[leftmargin=*,itemsep=2pt,topsep=4pt]
\item \emph{type 1:} both endpoints lie in $T$, and the edge is kept
with probability one;
\item \emph{type 2:} some endpoint was removed by the load test with
true load at least $1-O(\gamma)$.
\end{itemize}
Select one such endpoint from each edge of type~2 to form
a set $\calL$ of at most $r$ distinct vertices. The preceding bounds and the exceptional load estimate
imply, for a suitable $1\le\bar\ell=1+O(\gamma)$,
\begin{equation}\label{eq:load-window}
    \sum_{v\in C}(\ell_v-\bar\ell)_+=O(\gamma r),\qquad
    \sum_{v\in \calL}(1-\ell_v)_+=O(\gamma r).
\end{equation}
In words: almost no vertex of $C$ has surviving load much above one,
and almost every vertex of $\calL$ has surviving load close to one.

\paragraph{Part 4: a matching among the sampled edges.}
Let $k_1$ be the number of surviving edges of type~1 and
$k_2=|\calL|$ the number of type~2. All $k_1$ edges of type~1 are
kept. The $k_2$ edges of type~2 have contributed distinct vertices
whose total incident load is at least $k_2-O(\gamma r)$
by~\eqref{eq:load-window}.

Fix the prefix and the unordered suffix before exposing the acceptance
coins. Among the suffix edges with both endpoints in $C$, ignore vertices with
$\ell_v>2$. Their total incident load is $O(\gamma r)$, since
$\bar\ell\le 3/2$ for small $\gamma$ and
$\ell_v\le4(\ell_v-\bar\ell)_+$ at such vertices. Removing their edges
subtracts at most twice this load from the sum on $\calL$. Thus the
remaining edges have total $x$-mass at least $(k_2-O(\gamma r))/2$,
and each vertex has incident mass at most two.

For the analysis, thin the retained edges of this graph independently
so that edge $e$ is marked with probability $x_e/8$. This is possible
since $x_e/8\le p_e$. Conditional on an edge being marked, the
probability of another mark at either endpoint is at most
$(2+2)/8=1/2$. The isolated marked edges form a matching. Combining
their expected number with the $k_1$ edges kept for certain gives
\[
   \mathbb{E}\bigl[\nu(\Jhat[F])\bigr]\ge
      \max\{k_1,(k_2-O(\gamma r))/32\}
      \ge (k_1+k_2-O(\gamma r))/33 .
\]
As $k_1+k_2\ge |M_J|-O(\gamma r)$, we may take $\beta=1/33$.

\paragraph{Part 5: concentration.}
Assign each suffix edge to one of its endpoints in the fixed vertex
cover $\VC$. The acceptance coins form at most $2r$ independent groups,
one per cover vertex. Changing a group at $v$ changes only edges incident to
$v$, so the matching number changes by at most one, since deleting $v$
gives the same graph in both outcomes. McDiarmid's bounded differences inequality gives
a downward deviation of $\gamma r$ with probability at most
$\exp(-\gamma^2 r)$. Taking $\gamma$ small enough proves the claim.
The algorithm stores the edges accepted with probability $p_e$ and the per-vertex arrays. As in
Section~\ref{sec:implementation}, we cap the stored graph at $O(n)$ edges.
\end{proof}

\section{Estimating local eligibility}\label{sec:estimation}
We prove the two estimates used in Section~\ref{sec:sampling}.
\begin{itemize}[leftmargin=*,itemsep=2pt,topsep=4pt]
\item Proposition~\ref{prop:marker-estimation} approximates
eligibility by counts of constant-size rooted subgraphs, which can be
estimated from an edge sample.
\item Proposition~\ref{prop:stage-estimation} controls the cumulative
error when the algorithm sums these estimates over a vertex's
candidate neighbors, across all stages.
\end{itemize}
Both bounds scale with the matching number $r$, even if the graph has
many isolated vertices.

\paragraph{Setting.}
Retain the notation of the sampling rule. The reference graph $H$ has at
most $Ar$ edges, $X$ is the set of endpoints of its initial $p$-sample,
and $H'=H-X$, with $H_0$ obtained by the subsequent degree cutoff. Write $b_v\in\{0,1\}$ for
the output of the radius-$R$ eligibility
rule at $v$ on $H_0$, so that $F=\{v:b_v=1\}$. We write $d_H(v)$ for
the degree of $v$ in $H$. In the following estimates the failure
probability can be made at most $\delta$ by taking $r$ above a
sufficiently large fixed polynomial in $\Lambda=1+\log(2n/\delta)$.
We refer to this as \emph{the stated confidence}.

\subsection{Approximating eligibility by subgraph counts}

\paragraph{A warm-up.}
On a graph of maximum degree $d$, the indicator that a vertex of
degree $k$ is isolated equals $\sum_{j=0}^d(-1)^j\binom{k}{j}$.
Each term counts stars. From an edge sample of rate $q$ it has an
unbiased estimate by weighting each observed star by $q^{-j}$.
A general local rule uses other small connected graphs as well, and
the degree cutoff defining $H_0$ must be tested on $H'$. The initial
sample makes high degrees sufficiently rare that a polynomial can
approximate the cutoff with small total error.

We start with the consequences of the initial sample. A vertex of
high degree in $H$ is unlikely to avoid $X$, so every polynomially
growing function of the degree has a small sum over the surviving
vertices.

\begin{lemma}[Degree bounds after endpoint deletion]\label{lem:est-censor}
For a fixed nonnegative, polynomially bounded function $f$ and any $\zeta>0$,
with the stated confidence,
\begin{equation}
 \sum_{v\notin X:\,d_H(v)>0} f(d_H(v))
 \le 2Ar\sup_{k\ge1}(1-p)^k f(k)+\zeta r,
 \qquad \max_v d_{H'}(v)=O_p(\Lambda).                         \label{eq:est-censor}
\end{equation}
For fixed $s$ and $M$, the same confidence gives
$\sum_v(|\Ball_s^{H'}(v)|-1)^2=O(r)$ and at most $O(rD^{-M})$ vertices
within distance $s$ of a vertex whose degree in $H'$ exceeds $D$, for every
$D\ge1$.
\end{lemma}
\begin{proof}
A vertex of original degree $k$ survives with probability $(1-p)^k$,
and at most $2Ar$ vertices are nonisolated. This gives the stated
expectation bound. No vertex of degree above $C_p\Lambda$ survives except with
probability $ne^{-pC_p\Lambda}$. Omit such degrees for analysis.
Changing one of $Ar$ independent indicators affects at most two
terms, each of size $\poly(\Lambda)$. Bounded
differences gives the additive $\zeta r$ error.

For a connected configuration on $j$ vertices in $H'$, enumerate each
image from a vertex $w$ of largest original degree $k$. Its spanning
tree has at most $(1+k)^{j-1}$ images for that choice of $w$ and its
role. Pairs of nonconstant walks of length at most $s$ from a common
root bound the squared ball sizes and have finitely many such connected
types. Apply \eqref{eq:est-censor} to the resulting polynomial in $k$.
Similarly, a walk reaching a vertex of degree above $D$ has $k>D$.
Using a moment $M$ orders higher gives the factor $D^{-M}$, including
for walks of length zero at a high-degree vertex.
\end{proof}

\begin{proposition}[Approximating local rules]\label{prop:marker-estimation}
For every $\gamma>0$ there is a nonnegative $h_v$ given by a fixed linear
combination of connected rooted subgraph counts on $H'$, with bounded
coefficients computable from the labels of each copy and $h_v=1$ at isolates.
Let $c_v=\mathbf1_{\{b_v=1,\ h_v\ge1/2\}}$, $C=\{v:c_v=1\}$, and
$U=V\setminus C$. With the stated confidence,
\begin{align}
 \sum_v|h_v-b_v|&\le\gamma r,& \sum_v|h_v-c_v|&\le3\gamma r,\notag\\[-2pt]
 \sum_v|h_v-c_v|^2&=O(r),& |U|&\le(2A+2\gamma)r .             \label{eq:est-surrogate}
\end{align}
At most $2\gamma r$ accepted vertices lie outside $C$.
\end{proposition}
\begin{proof}
The proof has three steps: an exact expansion of the rule on $H_0$
into subgraph counts, a polynomial approximation of the degree cutoff
that moves the expansion to $H'$, and an error analysis.

\emph{Step 1: expansion on $H_0$.}
For an edge set $S$, let $b_v(S)$ be the rule's answer on the labeled
graph with edge set $S$ and root $v$. By inclusion--exclusion,
$b_v(H_0)$ is the sum over $E'\subseteq E(H_0)$ of
$\sum_{S\subseteq E'}(-1)^{|E'|-|S|}b_v(S)$. An edge outside the radius-$R$ ball
about $v$ in $E'$ cannot affect
the rule on any subset of $E'$. Pairing terms that differ only on this
edge cancels its coefficient.
Consequently the expansion on $H_0$ has constant one and finitely many
connected rooted support types of bounded size. On $H'$, count each
such labeled image with the additional factor
$\prod_{w\in V(E')}\mathbf1_{\{d_{H'}(w)\le d\}}$. This counts exactly
the images in $H_0$.

\emph{Step 2: a polynomial for the degree cutoff.}
For fixed $\vartheta>0$ there is a rational polynomial $\cutpoly$ approximating
$\mathbf1_{\{k\le d\}}$ with error at most $\xi e^{\vartheta k}$ at all integers
$k\ge0$, for arbitrarily small $\xi$. For completeness we give a
construction. Let $T_m$ denote the Chebyshev polynomial of degree
$m$, set $t=3+\lceil16m/\vartheta\rceil$, and use
\[
 Q_m(k)=T_m(1-2(k-1)/(t-1))/T_m(1+2/(t-1)).
\]
It equals one at zero, and has magnitude $O(e^{-\Omega(m/\sqrt t)})$
on $[1,t]$ by $T_m(\cosh x)=\cosh(mx)$. For $k\ge t$,
$e^{-\vartheta k/2}|Q_m(k)|\le e^{-\vartheta k/2}(8k/t)^m\le8^m e^{-\vartheta t/2}$. Thus
$\cutpoly(k)=\sum_{j=0}^d\binom{k}{j}Q_m(k-j)$ gives the claim for large
$m$, since $e^{-\vartheta k/2}\binom{k}{j}$ is bounded and the binomial vanishes
at integers $0\le k<j$.

\emph{Step 3: the score and its error.}
Replace the cutoff factors by $\cutpoly(d_{H'}(w))$ and square the resulting
sum to obtain $h_v$. Degree powers count ordered choices of incident
edges. Grouping coincidences gives fixed connected subgraph counts,
including in products with a common root. To compare with $b_v^2=b_v$,
let $j_0$ bound the number of vertices in an original support. Each
product uses at most $2j_0$ cutoff factors. For an image whose largest
original degree is $k$, telescoping the factors bounds its error by
$O(\xi e^{2j_0\vartheta k})$ for $\xi\le1$. The error also has a polynomial
bound in $k$ for fixed $\cutpoly$. Assign the image to a surviving vertex of
original degree $k$ as in Lemma~\ref{lem:est-censor}. Its total charge is bounded
by the deterministic function
$f(k)=\min\{R_\xi(k),C(1+k)^{2j_0}\xi e^{2j_0\vartheta k}\}$ for a polynomial
$R_\xi$ and a constant $C$ independent of $\xi$. Take $2j_0\vartheta<p/2$. Then
$\sup_k(1-p)^k f(k)=O(\xi)$, and \eqref{eq:est-censor} proves the
first error bound by choosing $\xi,\zeta$ small. The same enumeration
applied to the absolute coefficients of $h_v^2$ bounds its sum on
nonisolates by $O(r)$, using a fixed degree moment. At isolates
$h_v=b_v=c_v=1$. The remaining bounds follow since at most $2Ar$
vertices are nonisolated and $b_v=1,c_v=0$ costs at least $1/2$ in
the first error bound.
\end{proof}

\subsection{Controlling cumulative loads}
Recall from Section~\ref{sec:sampling} that for stage $i$ of the
sampling rule $\deg_i(v)=|N_i(v)|$,
$g_i(v)=\sum_{w\in N_i(v)}h_w$, and
\[
 L_v=\sum_{i<s}a_i g_i(v),\qquad
 L_v^c=\sum_{i<s}a_i |N_i(v)\cap C| .
\]
Both sums include the stage at which $v$ is removed. Fix
$0<\theta\le\gamma<1/16$ and take $A_0$ sufficiently large in
terms of $A,\theta,\gamma$.
\begin{proposition}[Degree and load estimates]\label{prop:stage-estimation}
With the stated confidence there is a set $\calE$ of \emph{exceptional
vertices} with $|\calE|\le\gamma r$ and
$\sum_{v\in \calE}L_v^c\le\gamma r$. Elsewhere $L_v\le1+O(\gamma)$ and
$L_v^c\le2L_v$. Every $v\notin \calE$ removed by its cumulative estimate
satisfies $L_v\ge1-O(\gamma)$. In addition,
\begin{align}
 \sum_{v\in C}|L_v-L_v^c|&=O(\gamma r),                       \label{eq:est-transfer}\\
 \sum_{v:\deg_i(v)>D_i}\deg_i(v)&\le rD_i^{-3},\qquad
 |E(J_{\rm remaining}[T_i])|=O(rD_i)\quad(0\le i\le s).       \label{eq:est-size}
\end{align}
Only $O(\gamma r)$ vertices of $C$ are removed by the degree test.
\end{proposition}

\paragraph{Proof outline.}
We first show a generic concentration statement for one degree scale.
We then prove, by induction over the stages, that few active vertices
have degree above the current scale, and that at each stage the
estimated load increment is accurate outside a small set $\calE_i$.
Finally we sum the errors over the stages; the geometric decay of the
scales makes the sums converge.

\begin{proof}
Condition on an initial reference-edge sample satisfying the preceding
bounds. Define the \emph{analysis weight} of a vertex by
$\wt(v)=1+|h_v-c_v|$.
Any $xr$ vertices have total analysis weight at most $(x+O(\sqrt x))r$ by
\eqref{eq:est-surrogate}, and $\max_v \wt(v)=\poly(\Lambda)$
by the maximum degree in \eqref{eq:est-censor}.

\paragraph{Step 1: concentration at one scale.}
Condition on the history before the
intervals under consideration. At scale $D=D_i$ the samples use at
most $Ar$ independent marked-edge indicators and $O(rD)$ independent
candidate-edge indicators, by the edge-count induction below. Suppose each
of $O(rD)$ events has probability at most $2e^{-\sqrt D}$, and changing
one indicator can affect at most $D^{O(1)}$ events. Give the events
fixed nonnegative weights with expected sum on occurring events at
most $rD^{-12}/2$. With the stated confidence their total weight is
at most $rD^{-12}$, provided each weight is $\poly(\Lambda)$
when $\sqrt D\le C\Lambda$. Above this scale a union bound suffices.
Below it, changing one indicator changes the sum by
$\poly(\Lambda)$ and there are only
$r\poly(\Lambda)$ indicators. Bounded differences at
deviation $rD^{-12}/2$ gives failure
$\exp[-\Omega(r/\poly(\Lambda))]$.

\paragraph{Step 2: remaining degrees.}
Inductively suppose there are $O(rD)$
remaining edges and strengthen \eqref{eq:est-size} to
\begin{equation}
            \sum_{v:\deg_i(v)>D}\wt(v) \deg_i(v)\le rD^{-12}.          \label{eq:est-large-degrees}
\end{equation}
Both hold at $i=0$. Before the third interval only $O(r)$ vertices
have degree $k>D_{i+1}$, and their total analysis weight is $O(r)$. Each
survives with probability $e^{-\Omega(q_i k)}$ by the binomial tail.
The expected surviving weight, charging $\wt(v)k$ per vertex, is at most
$O(r)\sup_{k>D_{i+1}}k e^{-\Omega(q_i k)}$. This meets the preceding
concentration criterion at degree $D_{i+1}$ for large $D_*$. Each
sampled edge affects two degrees. Degrees above $C\Lambda/q_i$ may
be omitted for analysis because a union bound rules out any such survivor with the stated
confidence. At smaller scales all remaining weights are polynomial in $\Lambda$.
Remaining degrees only decrease, proving \eqref{eq:est-large-degrees}
at $i+1$. All edges avoiding these high degrees meet the fixed cover
of size $2r$ and hence number at most $2rD_{i+1}$. This gives the
edge-count induction with a uniform constant. In the other direction,
a positive degree below $D/4$ fails the third test with probability
$e^{-\Omega(q_iD)}$. There are $O(rD)$ such vertices and again each
edge affects two tests, so at most $rD^{-12}$ are wrongly removed.

\paragraph{Step 3: accuracy of each increment.}
We compare with truncated graphs on which all quantities are bounded.
For analysis delete from $H'$ the
vertices of degree above $D^\psi$, for a small fixed $\psi>0$, and delete
from the active candidate graph the vertices of degree above $D$. Use the
same inclusion indicators on every edge shared with the original
graphs. A sampled score can change only within the fixed
subgraph radius of a vertex deleted from the reference graph. By Lemma~\ref{lem:est-censor}
there are at most $C_MrD^{-\psi M}$ such roots for any fixed $M$. Deleted
vertices of the candidate graph and their neighbors number at most $2rD^{-12}$ by
\eqref{eq:est-large-degrees}. Each remaining exceptional root has
at most $D$ candidate neighbors. Outside these $O(rD^{-12})$ vertices of the candidate graph
(take $\psi M>14$), the neighbor list and every sampled score on it agree
in the two graphs. Thus both $\hg_i(v)$ and its true target $g_i(v)$
agree, for every realization of the common indicators.

On the restricted graphs a score $\hh_i(w)$ has magnitude at most
$O(D^{\psi K}q_i^{-K})$ for a constant $K$ depending on the subgraph
types. Increasing
$K$ if needed, its neighborhood contains $O(D^{\psi K})$ marked edges,
and any marked edge lies in the neighborhoods of $O(D^{\psi K})$ roots.
The estimate at $v$ is unbiased because the two samples
use disjoint label sets. Its sum over at most $D$ neighbors depends
on $O(D^{1+\psi K})$ independent indicators. Changing one marked-edge
indicator changes at most $O(D^{\psi K})$ summands, each of magnitude
$O(D^{\psi K+\phi(K+1)})$ after division by $q_i$. Changing a candidate-edge
indicator changes at most one summand. Bounded differences at unscaled
deviation $D^{-\chi}/a_i$ therefore gives
\[
 \Pr\{|a_i(\hg_i(v)-\mathbb{E}[\hg_i(v)])|>D^{-\chi}\}
 \le 2\exp[-\Omega(D^{1-2\chi-5\psi K-2\phi(K+1)})]\le 2e^{-\sqrt D}
\]
for sufficiently small positive $\psi,\phi,\chi$ and large $D_*$. Across all
vertices of the candidate graph a marked-edge indicator affects at most $O(D^{1+\psi K})$
estimates, and a candidate-edge indicator at most two. Zero degrees incur
no error. The concentration bound of Step~1 therefore gives at most
$rD^{-12}$ additional failures. Together with the excluded vertices
and incorrect degree removals this gives a set $\calE_i$ of total
analysis weight at most $rD^{-4}$ by \eqref{eq:est-surrogate}. Outside
$\calE_i$ the true increment differs from $a_i\,\hg_i(v)$ by at most $D^{-\chi}$.

\paragraph{Step 4: accumulating the loads.}
Put $\calE=\bigcup_i \calE_i$. Until its first
exception a vertex has total estimation error at most
$\sum_i D_i^{-\chi}=O_{\chi,\theta}(D_*^{-\chi})$. Since $h\ge0$, its neighbor
sum is nonincreasing. The identities $a_i=a_0+\theta\sum_{j<i}a_j$
and $a_0\sum_w h_w\le\theta$ (since $\sum_w h_w\le n+\gamma r$) give
$a_i g_i(v)\le\theta(1+\sum_{j<i}a_j g_j(v))$. Before removal the
earlier load is at most one plus the estimation error. This proves
both asserted threshold bounds outside $\calE$ for large $D_*$.

If a vertex first fails at scale $D_i$, its earlier eligible load is
at most twice its true load since $c_v\le2h_v$. There are only
$O_\theta(1+\log D_i)$ stages left. At degree at most $D_j$ each costs
at most $a_jD_j=\theta A_0$ in eligible load. Larger degrees are
already bounded in \eqref{eq:est-large-degrees}. Therefore
\[
 \sum_{v\in \calE}\wt(v)(1+L_v^c)
 \le O(r)\sum_i D_i^{-4}(1+\log D_i)+r\sum_i a_iD_i^{-12}
 \le\gamma r .
\]
Exchanging the two endpoints of each active edge gives
$\sum_{v\in C}|L_v-L_v^c|\le\sum_w|h_w-c_w|L_w^c=O(\gamma r)$ by
this bound on $\calE$ and \eqref{eq:est-surrogate} elsewhere.

\paragraph{Step 5: vertices of $C$ removed by the degree test.}
A vertex of $C\setminus \calE$ discarded by the third interval had true
stage-start eligible degree at most $2(1+O(\gamma))/a_i$ because it
passed the cumulative-estimate test. Its degree before the third
interval was at least $D_i/4$. Choosing $\theta A_0$ large gives at
least $D_i/8$ neighbors in $U$. Charge $a_i$ per such incidence.
Each discarded vertex costs at least $\theta A_0/8$, whereas the total
charge is at most $\sum_{u\in U}L_u^c=O_A(r)$. Taking $A_0$ large
enough bounds the number discarded by $\gamma r$, plus vertices of $\calE$.
Finally the conditional estimates at all $O(\log n)$ stages hold
simultaneously by a union bound using Section~\ref{sec:clocks}.
\end{proof}

\section{Arbitrary weights and unknown stream length}
\label{sec:implementation}
We complete the proof of Theorem~\ref{thm:main}.
Theorem~\ref{thm:bounded-general} stores a sparse subgraph containing a
near-optimal matching, under two assumptions: a known stream length
and integer weights in $[1,W]$. This section recovers a matching from
the stored subgraph (Section~\ref{sec:recover}), removes the two
assumptions (Sections~\ref{sec:length} and~\ref{sec:weights}), and
accounts for space, precision, and running time
(Section~\ref{sec:space}). Let $b$ be the smallest power of
two at least $20/\epsilon$ and put $\eta=1/b$.

\subsection{Recovering a matching}\label{sec:recover}
Use Theorem~\ref{thm:bounded-general} with error $\eta$. On the stored
integer-weight graph, repeatedly flip a short augmentation, that is, a
short alternating path or even
cycle that increases the matching weight. Lemma~\ref{lem:short-local-seed}
guarantees a feasible augmentation of length $O(W/\eta)$ until the
matching attains $(1-\eta)$ of the stored optimum. Enumerating all
such paths and cycles takes polynomial time for fixed $W,\eta$ and
requires only the edge list, the matching and one candidate at a time.
At most $Wn$ flips increase the integer weight.

\subsection{Unknown stream length}\label{sec:length}
For weights in $[1,W]$, put $a=1+\eta/W$. Run parallel copies for
the prefix-length guesses $N_0=1$ and $N_{i+1}=\lceil aN_i\rceil$,
for $0\le i\le\lceil\log_a(n^2)\rceil$. Cap each copy's horizon at $n^2$.
For an input of $m>0$ edges, select the largest guess $N\le m$.
It satisfies $m\le aN$. Fix an optimum matching $P$. The
omitted tail contains a hypergeometric number of its edges with mean at
most $\eta |P|/W$. Except with probability $\delta'$, the omitted tail
costs at most $2\eta w(P)+O(W\log(2/\delta'))$ in weight, by the
hypergeometric upper tail and the bound $W$ on each weight. The loss is
at most $3\eta w(P)$ for $w(P)\ge C W\log(2/\delta')/\eta$ with $C$
large enough. Below this threshold a parallel buffer of
$2CnW\log(2/\delta')/\eta$ edges stores the entire input, since
$|E(G)|\le2n\nu(G)\le2nw(P)$.

Conditional on the edge set of the prefix, its order is uniform, so the
fixed-length guarantee applies to its own optimum. If the input fits in the
buffer, decode it directly. Otherwise, return the matching from the largest
completed prefix. In either case its weight is at least
$(1-5\eta)w(P)$ after accounting for the prefix loss and the two
$(1-\eta)$ factors. An empty stream returns the empty matching.

\subsection{Removing the weight bound}\label{sec:weights}
We use the weight-class reduction of Gupta and Peng~\cite{gp} as in
random-order streaming~\cite{hw}.

\paragraph{Weight classes and blocks.}
Class $j\in\mathbb Z$ consists of the
edges with weights in $[b^j,b^{j+1})$. For each residue $i$ modulo $b$,
omit classes $j\equiv i\pmod b$. The remaining blocks of $b-1$
consecutive classes are separated by a factor of $b$ in weight.
For some choice of $i$, the omitted classes carry at most $w(P)/b$
of a fixed optimum matching $P$, by averaging.

\paragraph{Running the bounded-weight algorithm on a block.}
For a block with lower boundary $a$, round weights down to multiples of
$a/b$. Rounding preserves at least $(1-\eta)w_e$ on every edge, and the
weights in units of $a/b$ are integers at most $b^b$. Run the bounded-weight
algorithm on the block's edges in their original order, a uniform
permutation of that block. Keep the original weight with each retained
edge for the final comparison.

\paragraph{Bounding the number of live blocks.}
Track the greatest
weight $w_{\max}$ seen so far. Discard blocks with upper boundary at
most $\eta w_{\max}/n$ and ignore later edges in them. This monotone
cutoff leaves $O_\epsilon(\log(2n))$ nonempty blocks live per residue.
Each block live at the end has received its entire subsequence. The
weight of $P$ in discarded blocks is at most $\eta w_{\max}\le\eta w(P)$.

\paragraph{Merging the block matchings.}
For each residue, merge its block matchings in decreasing block order,
adding an edge whenever both endpoints are free. An added edge of
weight $w$ can block at most two edges in each lower block, one per
endpoint. In the $\ell$th lower block such an edge weighs at most
$w/b^\ell$. Thus the total rejected weight is at most
$2/(b-1)$ times the merged weight. For the residue identified by
averaging, the live blocks contain at least $(1-2\eta)w(P)$ of the
optimum. Returning the heaviest merged matching, using original weights,
therefore gives at least
\[
     \frac{(1-5\eta)(1-\eta)(1-2\eta)}{1+2/(b-1)}\,w(P)
          \ \ge\ (1-\epsilon)w(P).
\]

\subsection{Space, precision, and confidence}\label{sec:space}

\paragraph{Confidence.}
For each residue at most $\binom n2$ blocks ever contain an input edge.
Analyze the full subsequence of each block. The blocks live at the end
depend only on the unordered input. A union bound over all blocks,
$O_\epsilon(\log n)$ prefix lengths and $O_\epsilon(\log n)$ stages per
copy gives $\log(1/\delta')=O_\epsilon(\Lambda)$ for each failure event.
It suffices to specify confidence by $\lceil\log_2(1/\delta)\rceil$.

\paragraph{Space.}
Each copy stores a fixed number of subgraphs and per-vertex arrays.
Enforce a deterministic space bound by capping every $O(r)$ subgraph
at its bound with $n$ in place of $r$. The parallel full-input buffer
has capacity $O_\epsilon(n\Lambda^{C_\epsilon})$. An overflowing
copy is discarded. The empty matching is always an admissible output.

\paragraph{Precision.}
Choose rational fixed parameters and dyadic sampling rates within a
factor of two of the specified rates, respecting the equal lengths of
the first two intervals of each stage. Use the actual probabilities in
the estimates. Each relative interval length takes $O_\epsilon(\log n)$
bits. Within a stage, a subgraph count is polynomial in $n$ and its
inverse-probability weights are bounded powers of the same rational
number. Including the fixed coefficients and geometric stage multipliers,
the boundary times and cumulative loads over $O_\epsilon(\log n)$
stages have polylogarithmic bit length. Enumerate one rooted subgraph
at a time. The local price computation uses a fixed number of rational
steps.

\paragraph{Randomness and running time.}
A binomial draw uses at most $n^2$ rational coin flips.
Rejection sampling for a coin of bias $a/d$ draws a uniform integer
in $[d]$ with probability at least $1/2$ per attempt. Truncating
each of polynomially many draws after $O_\epsilon(\log(n/\delta))$
attempts gives polynomial worst-case time within the failure budget.
Each original edge uses $O(B+\log n)$ bits and each matching weight
can be summed exactly in $O(nB+\log n)$ bits. Summing over live
copies yields the space bound in Theorem~\ref{thm:main}.

\section*{AI use and formal verification}
AI models from OpenAI and Anthropic assisted with the literature
search, proof development and checking, writing, and Lean formalization.
Theorem~\ref{thm:main} is fully verified in Lean, including the
success probability, running time, and peak space of the one-pass algorithm.
The formal proofs are available at
\url{https://github.com/farhadi94/random-order-matching-ptas}.

\end{document}